\documentclass[aps,prl,reprint,superscriptaddress,twocolumn,longbibliography]{revtex4-1}

\usepackage{graphicx}
\usepackage{bm}
\usepackage{multirow}
\usepackage{color}
\usepackage{braket}
\usepackage{physics}
\usepackage{upgreek}
\usepackage[colorlinks, linkcolor=blue, urlcolor=blue, anchorcolor=blue, citecolor=blue]{hyperref}

\begin{document}


\title{Experimental realization of a 218-ion multi-qubit quantum memory}

\author{R. Yao}
\thanks{These authors contribute equally to this work}%
\affiliation{Center for Quantum Information, Institute for Interdisciplinary Information Sciences, Tsinghua University, Beijing 100084, PR China}

\author{W.-Q. Lian}
\thanks{These authors contribute equally to this work}%
\affiliation{HYQ Co., Ltd., Beijing 100176, PR China}

\author{Y.-K. Wu}
\affiliation{Center for Quantum Information, Institute for Interdisciplinary Information Sciences, Tsinghua University, Beijing 100084, PR China}
\affiliation{Hefei National Laboratory, Hefei 230088, PR China}

\author{G.-X. Wang}
\affiliation{Center for Quantum Information, Institute for Interdisciplinary Information Sciences, Tsinghua University, Beijing 100084, PR China}

\author{B.-W. Li}
\affiliation{Center for Quantum Information, Institute for Interdisciplinary Information Sciences, Tsinghua University, Beijing 100084, PR China}

\author{Q.-X. Mei}
\affiliation{HYQ Co., Ltd., Beijing 100176, PR China}

\author{B.-X. Qi}
\affiliation{Center for Quantum Information, Institute for Interdisciplinary Information Sciences, Tsinghua University, Beijing 100084, PR China}

\author{L. Yao}
\affiliation{HYQ Co., Ltd., Beijing 100176, PR China}

\author{Z.-C. Zhou}
\affiliation{Center for Quantum Information, Institute for Interdisciplinary Information Sciences, Tsinghua University, Beijing 100084, PR China}
\affiliation{Hefei National Laboratory, Hefei 230088, PR China}

\author{L. He}
\email{heli2015@mail.tsinghua.edu.cn}
\affiliation{Center for Quantum Information, Institute for Interdisciplinary Information Sciences, Tsinghua University, Beijing 100084, PR China}
\affiliation{Hefei National Laboratory, Hefei 230088, PR China}

\author{L.-M. Duan}
\email{lmduan@tsinghua.edu.cn}
\affiliation{Center for Quantum Information, Institute for Interdisciplinary Information Sciences, Tsinghua University, Beijing 100084, PR China}
\affiliation{Hefei National Laboratory, Hefei 230088, PR China}

\date{\today}

\begin{abstract}
Storage lifetime and capacity are two important factors to characterize the performance of a quantum memory. Here we report the stable trapping of above $200$ ions in a cryogenic setup, and demonstrate the combination of the multi-qubit capacity and long storage lifetime by measuring the coherence time of randomly chosen ions to be on the order of hundreds of milliseconds. We apply composite microwave pulses to manipulate qubit states globally for efficient characterization of different storage units simultaneously, and we compare the performance of the quantum memory with and without the sympathetic cooling laser, thus unambiguously show the necessity of sympathetic cooling for the long-time storage of multiple ionic qubits.
\end{abstract}

\maketitle

The quantum memory is an important building block in quantum technology \cite{lvovsky2009optical}. For long-distance quantum communication and quantum cryptography, it lies at the core of the quantum repeater protocol which has an exponential improvement in the communication efficiency \cite{PhysRevLett.81.5932,duan2001long,RevModPhys.83.33}. For quantum computation, it synchronizes the qubits by appending identity gates between different quantum operations, and it allows the preparation of ancilla states in advance, which comprises the major cost of the fault-tolerant quantum computing \cite{Gottesman1998,campbell2017roads,PhysRevA.86.032324}. Furthermore, matter qubits like trapped ions \cite{doi:10.1063/1.5088164} and superconducting circuits \cite{doi:10.1146/annurev-conmatphys-031119-050605,huang2020superconducting} by themselves can be regarded as quantum memories, whose performance fundamentally bounds those of all the quantum operations on these qubits.

Several figures of merit are used to characterize a quantum memory, such as its storage fidelity, lifetime and capacity \cite{lvovsky2009optical}. For applications in quantum networks, conversion fidelity and efficiency between matter qubits and photonic qubits are also concerned. For example, atomic ensembles have demonstrated single-excitation storage lifetime around $0.2\,$s \cite{yang2016efficient}, single-qubit storage fidelity of 99\% and efficiency of 85\% \cite{wang2019efficient}, and storage capacity of 105 qubits \cite{jiang2019experimental} respectively in individual experiments; solid-state spins based on rare-earth-doped crystals have achieved storage lifetime of tens of milliseconds \cite{ortu2022storage} and the capacity of tens of temporal modes \cite{lago2021telecom}, and a spin ensemble coherence time above 6 hours \cite{zhong2015optically}; the NV center in a diamond \cite{PhysRevX.9.031045} and the neutral atoms in optical traps \cite{barnes2022assembly} have also realized respectively ten or tens of qubit storage for a lifetime of about one minute. As one of the leading quantum information processing platforms, trapped ions keep the record for the longest single-qubit storage lifetime about one hour \cite{wang2017single,wang2021single}. Entanglement between ionic and photonic qubits has also been demonstrated \cite{blinov2004observation,stute2012tunable,hucul2015modular,bock2018high,krutyanskiy2019light,PhysRevLett.124.110501} as a plausible way to scale up the ion trap quantum computer \cite{10.5555/2011617.2011618,duan2010colloquium,hucul2015modular}. However, the long storage lifetime reported in Refs.~\cite{wang2017single,wang2021single} for a single qubit cannot be directly extended to the multi-qubit quantum memory case due to many technical challenges. For instance, a long ion chain is subjected to stronger background gas collisions and motional heating from the electric field noise \cite{wineland1998experimental,doi:10.1063/1.5088164} than a few ions, which corrupts the stability of the ion chain and causes difficulty in the readout of the qubit states. Besides, even if the ion crystal is maintained, the frequent thermal hopping of the ions in a room-temperature trap will still randomize the location of the stored qubits and thus destroy the multi-qubit memory.
To date, above $100$ ions have been trapped in a crystal with spatial resolution in a cryogenic setup \cite{Pagano_2018}, and the quantum simulation operation and the individual readout of above $60$ ions have been achieved in a room-temperature trap \cite{2208.03060}. Shallow-depth quantum circuits composed of high-fidelity single-qubit and two-qubit gates have been realized for tens of ions \cite{egan2021fault,postler2022demonstration,2208.01863}, which provide an upper bound on the noise for a storage time of about milliseconds. However, longer storage lifetime for the multi-qubit system has not been demonstrated yet.

In this paper, we make use of a cryogenic Paul trap to maintain the stability of long ion chains and quasi-1D zigzag crystals, up to $218$ ions, by sympathetically cooling a small fraction of ions in the middle \cite{PhysRevLett.127.143201}, and demonstrate that arbitrarily chosen storage ions on the edges can achieve a typical storage coherence time above hundreds of milliseconds. This represents the largest number of ion qubits reported so far that can be stably stored in a Paul trap with sufficiently long quantum coherence time. Our work thus showcases that the multi-qubit storage capacity and the long quantum coherence time can be combined together for the trapped ion crystals, which can find applications in quantum computing, quantum networks, and other quantum information protocols where both the storage capacity and the long coherence time are required for the associated quantum memory.

\begin{figure*}[!tbp]
	\centering
	\includegraphics[width=\linewidth]{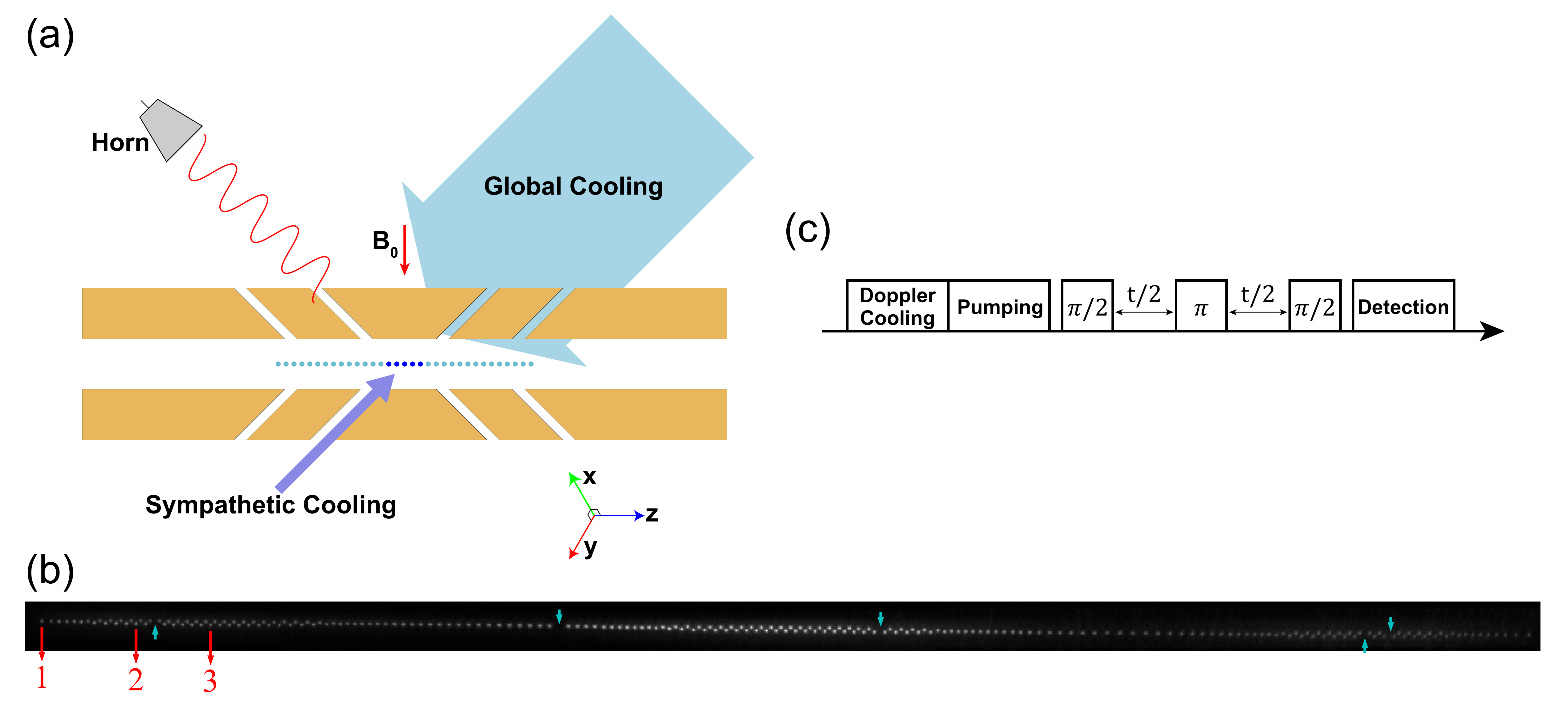}
	\caption{(a) Schematic experimental setup. A 1D or quasi-1D ion crystal is confined in a blade trap under cryogenic temperature. A broad elliptic beam with tunable frequency and intensity can be used for global Doppler cooling, optical pumping and qubit state detection, and a narrow sympathetic cooling beam propagating in the opposite direction can address about 5 central ions. A microwave horn antenna generates microwave signals to manipulate the states of the qubits. (b) The image of 218 ions in a zigzag structure. The length of this quasi-1D crystal is about $800\,\upmu$m and is wider than the FOV of the EMCCD camera of about $300\,\upmu$m, so the image is stitched from three images taken sequentially with overlap in between. The blue arrows indicate the locations of five dark ions. The ions labelled as 1, 2 and 3 are used to demonstrate quantum storage in the following experiments. (c) Experimental sequence. We initialize the qubit state through global Doppler cooling and optical pumping, and then use a global microwave $\pi/2$ pulse to prepare the qubit into the desired state. After the long-time storage with a spin echo (microwave $\pi$ pulse) in the middle, we apply another microwave pulse to reverse the preparation step and finally measure the storage fidelity of the qubit state. Due to the nonuniformity of the microwave, we use SK1 composite pulses for the $\pi/2$ and $\pi$ pulses to suppress the pulse area error. }\label{fig:setup}
\end{figure*}

To get a stable long ion chain or quasi-1D ion crystal, we use a blade trap in a closed-cycle cryostat \cite{Pagano_2018} at a temperature of $6\,$K. As shown in Fig.~\ref{fig:setup}(a), we apply a global Doppler cooling beam with nonzero angles to all the three principle axes. This global beam has $20\,\upmu$W power and $\Delta=-2\pi\times 12\,$MHz detuning, and is shaped into an ellipse with waist sizes of $15\,\upmu$m$\times 500\,\upmu$m by a cylindrical lens. Opposite to this beam is a narrow $1\,\upmu$W sympathetic cooling beam with the same detuning, which is focused to a beam waist of $10\,\upmu$m to address about 5 ions in the center of the crystal. By setting the transverse trap frequencies $\omega_x\approx 1.6\,$MHz, $\omega_y\approx 1.5\,$MHz, and by engineering the axial trapping potential via the segmented electrodes (for the long ion crystal, the axial potential cannot be well approximated by a harmonic trap), we obtain a quasi-1D ion crystal of 218 $^{171}\mathrm{Yb}^+$ ions in a zigzag shape, which can be stably trapped for hours under the global cooling beam.

An image of the $218$ ions is shown in Fig.~\ref{fig:setup}(b). The whole quasi-1D crystal spans a length of $800\,\upmu$m, which is beyond the field of view (FOV) of the EMCCD camera. For an illustration of the whole ion crystal, we take photos sequentially for different parts of the ion crystal by shifting the position of the camera, and perform image stitching to combine them together. In the following when characterizing the storage capacity and the lifetime of the quantum memory, we apply sympathetic cooling laser to the middle ions \cite{PhysRevLett.127.143201} and store and read out qubit states on the edge, so that the shifting of the camera position is not needed.

We apply a global microwave to manipulate the qubit states through the spin-echo sequence shown in Fig.~\ref{fig:setup}(c): We initialize the storage qubit in $|0\rangle \equiv |^{2}\mathrm{S}_{1/2},F=0,m_F=0\rangle$ of the $^{171}\mathrm{Yb}^+$ ion (a dark state under $370\,$nm detection laser) through optical pumping, and apply a $\pi/2$ pulse to prepare it in $(|0\rangle+e^{i\phi}|1\rangle)/\sqrt{2}$ where $\phi$ is the initial phase of the microwave signal. After a storage time $t$ with a $\pi$ pulse in the middle, we apply another $\pi/2$ pulse with opposite phase to reverse the state preparation process. Ideally we will end in the state $|1\rangle \equiv |^{2}\mathrm{S}_{1/2},F=1,m_F=0\rangle$ (a bright state under the detection laser), and the decay in the population versus time can give us the storage lifetime of the quantum memory.

\begin{figure*}[!tbp]
	\centering
	\includegraphics[width=\linewidth]{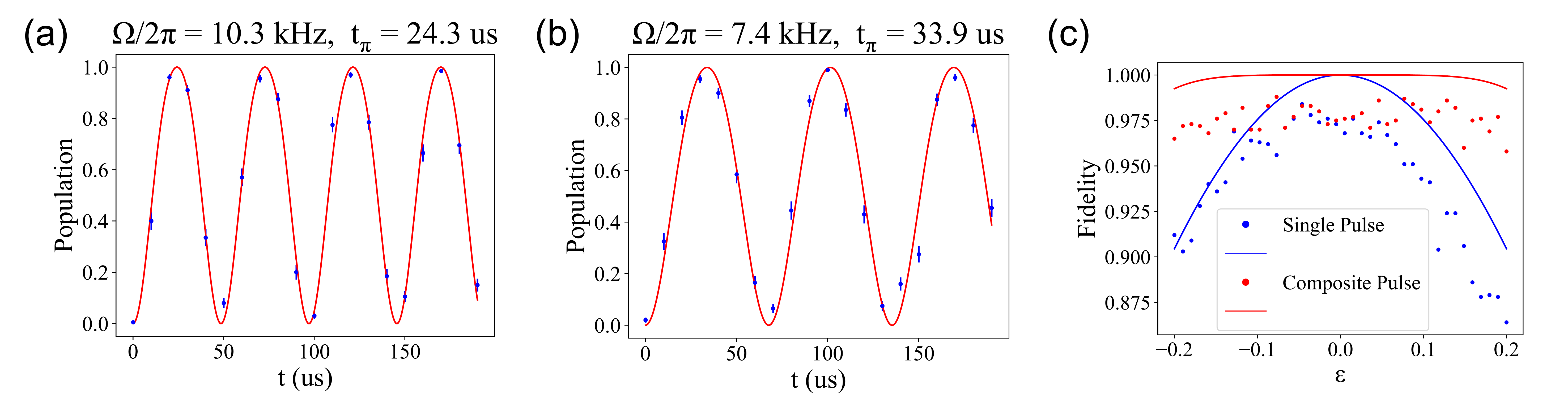}
	\caption{(a), (b) Here we use the Rabi oscillation of different sites in a chain of about 100 ions to characterize the nonuniformity of the microwave signal. For two ions separated by 80 ion spacings, the Rabi frequencies of (a) $2\pi\times 10.3\,$kHz and (b) $2\pi\times 7.4\,$kHz are fitted. (c) Theoretical (solid curves) and experimental (dots) performance for the single pulse (blue) and the composite pulse (red) on a target ion. Theoretically, the SK1 composite pulse can tolerate up to $\pm 20\%$ pulse-area error while still maintaining a fidelity above $99\%$. Similar tendency is observed in the experiment and the overall reduction in the fidelity can be explained by the state-preparation-and-measurement (SPAM) error.}\label{fig:composite}
\end{figure*}

Due to the nonuniformity of the microwave signal, its Rabi frequency varies slightly for different ions, so we use the SK1 composite pulse \cite{PhysRevA.70.052318,doi:10.7566/JPSJ.82.014004} to suppress this pulse-area error to higher order. As shown in Fig.~\ref{fig:composite}, there is about $30\%$-$40\%$ change in the Rabi frequency over 80 ion spacings, or about $0.5\%$ change between adjacent ions, and the SK1 pulse can tolerate up to $\pm20\%$ error in the Rabi frequency while still maintaining a fidelity above $99\%$. Therefore, we can simultaneously initialize all the ions in the FOV of the CCD camera in the same state and operate them with the same gate, which allows efficient characterization of their storage lifetime without the need to repeat the experimental sequence for each ion. Note that here our purpose is to demonstrate the storage capacity and lifetime of individual ions in the crystal, so global operations plus individual detection suffice. In the future with an upgrade in the addressing system using focused Raman laser beams, the simultaneous storage of multiple qubits into nearby ions can also be achieved.

\begin{figure}[!tbp]
	\centering
	\includegraphics[width=\linewidth]{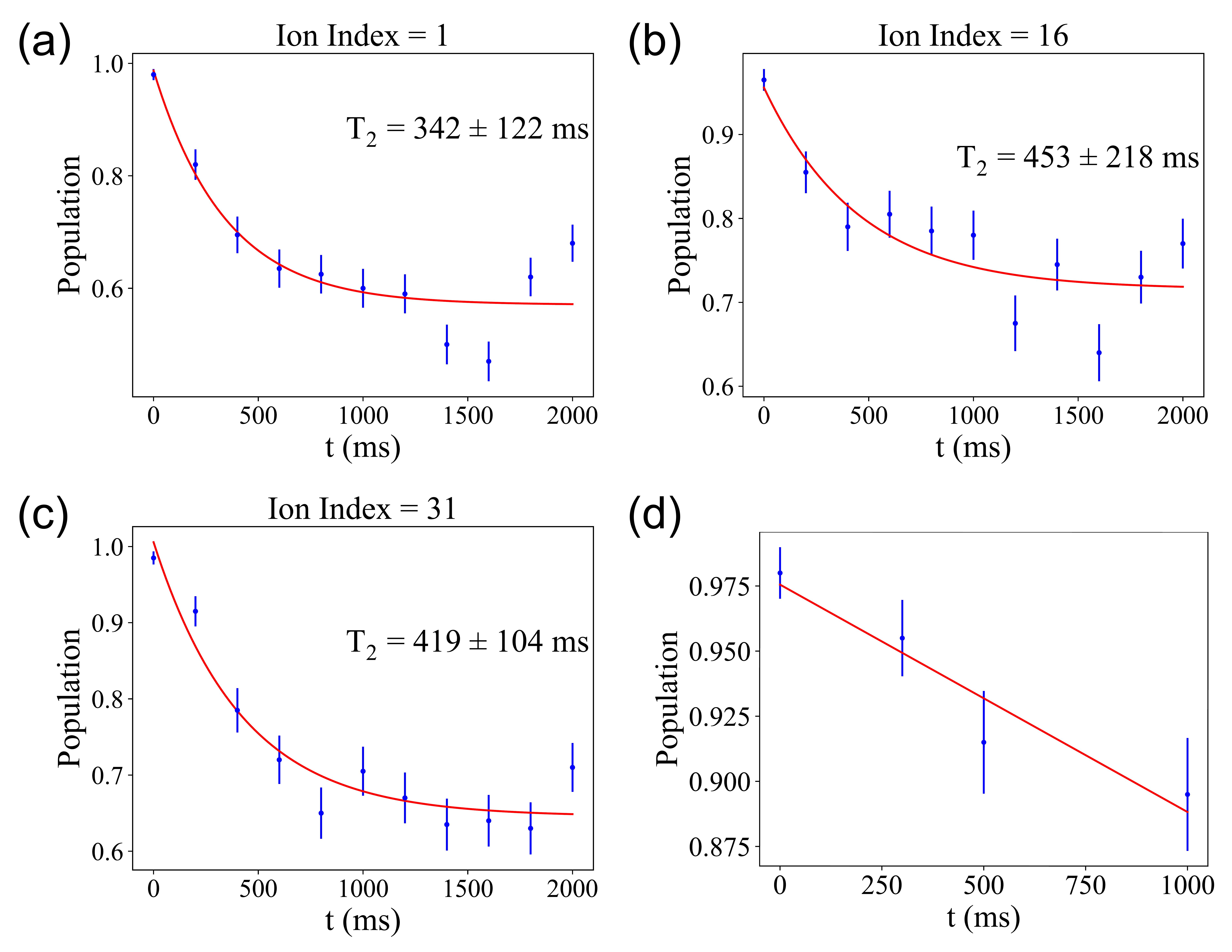}
	\caption{(a)-(c) The measured storage fidelity vs. storage time for three typical ions on the left side of the quasi-1D ion crystal in Fig.~\ref{fig:setup}(b). Here we measure the average fidelity in the $|+\rangle$, $|-\rangle$, $|L\rangle$, $|R\rangle$ bases using the spin-echo sequence in Fig.~\ref{fig:setup}(c). The coherence time $T_2$ is fitted to be $(323\pm 93)\,$ms, $(453\pm218)\,$ms and $(419\pm104)\,$ms, respectively, by the exponential function $F=A+Be^{-t/T_2}$. Each data point is repeated for 200 times with the error bar indicating one standard deviation. At large $t$, some data points deviate considerably from the exponential fitting function. This can be explained by the slow drift in the trap and laser parameters on the timescale of minutes to hours, and should not affect the extracted coherence time of hundreds of milliseconds. (d) The decay of the bright state ($|1\rangle$) population vs. storage time without the spin echo. A much longer relaxation time beyond $1\,$s is observed.}\label{fig:storage}
\end{figure}

We measure the storage lifetime of three typical ions [labeled as 1-3 in Fig.~\ref{fig:setup}(b)] away from the center of the ion crystal in Fig.~\ref{fig:storage}(a)-(c). By executing the pulse sequence in Fig.~\ref{fig:setup}(c), we get the average storage fidelity over the $|\pm\rangle=(|0\rangle\pm|1\rangle)/\sqrt{2}$ and $|L(R)\rangle=(|0\rangle\pm i|1\rangle)/\sqrt{2}$ bases \cite{wang2021single}, and fit the decoherence time $T_2=(323\pm 93)\,$ms, $(453\pm218)\,$ms and $(419\pm104)\,$ms, respectively, for the three ions. The relatively large error bar is mainly caused by the slow drift in trap and laser parameters on the timescale of minutes to hours, thus deviation from the exponential fit. Nevertheless, a coherence time on the order of hundreds of milliseconds can be concluded. On the other hand, the $|0(1)\rangle$ basis is not subjected to the dephasing error and hence typically has longer coherence time \cite{wang2021single}. For example, in Fig.~\ref{fig:storage}(d) we see that, without the spin echo in the middle which does not affect the decoherence in the $|0(1)\rangle$ basis, the population decay from the bright state ($|1\rangle$) to the dark state ($|0\rangle$) for a typical ion has a timescale far above $1\,$s. Therefore we can bound the average storage lifetime over all possible qubit states by $T_2$ for individual ions.
\begin{figure*}[!tbp]
	\centering
	\includegraphics[width=\linewidth]{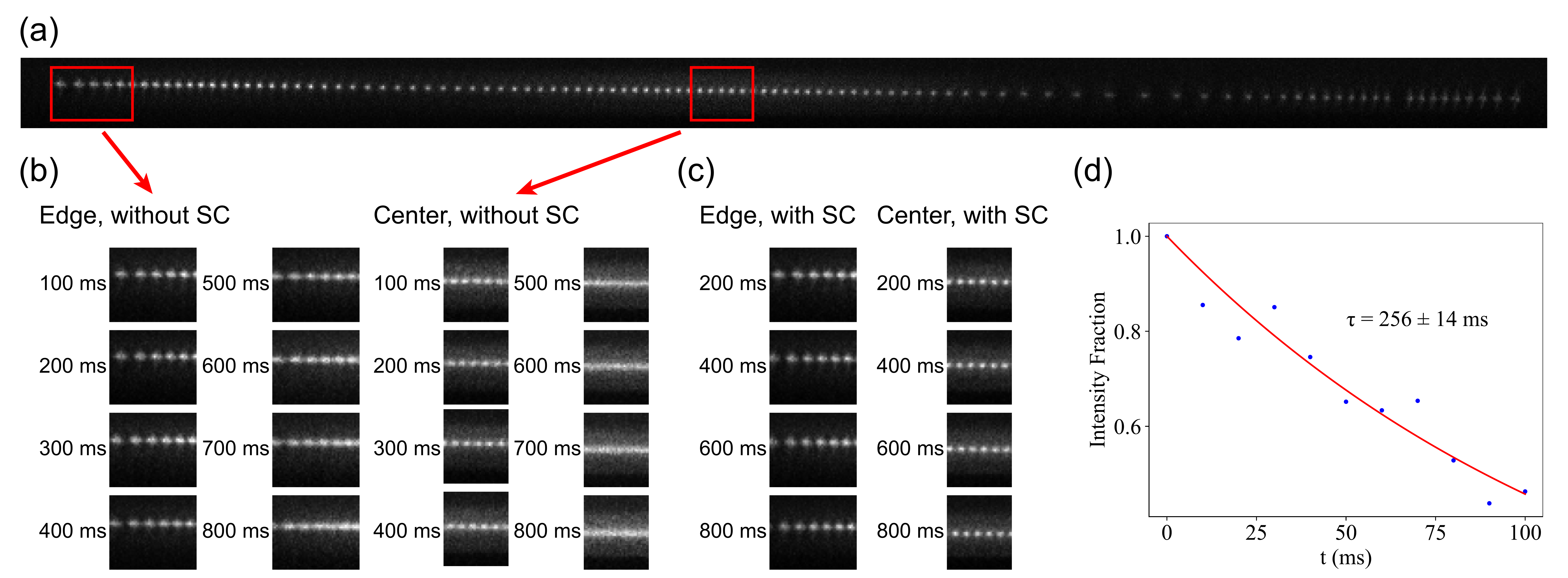}
	\caption{(a) The image of a 1D 103-ion chain under the same parameters as those for the crystal in Fig.~\ref{fig:setup}(b). To compare the heating effect with or without sympathetic cooling, we start from the same equilibrium state under global Doppler cooling, and then turn off the global cooling beam and evolve the system with the sympathetic cooling (SC) beam off or on.  Each image is averaged over 200 repetitions. (b) The images of the edge ions and the central ions [indicated by the red boxes in (a)] with the sympathetic cooling beam turned off. As the heating dynamics proceeds, the spots of individual ions dim and expand, making it difficult to detect individual qubit states. (c) The images of the same edge ions and the same central ions with the sympathetic cooling beam turned on. There is no visible change in these images after $800\,$ms evolution. (d) The decay of photon counts in a $5\,$pixel$\times5\,$pixel box surrounding a typical ion in the 218-ion crystal. Again, the photon counts quickly shrink with time, reducing the readout fidelity of the stored qubits.}\label{fig:cooling}
\end{figure*}

The sympathetic cooling laser beam turns out to be critical for the functioning of this quantum memory: without the cooling beam, the ion crystal can quickly be heated up within hundreds of milliseconds that can cause difficulty in reading out the qubit state or even the meltdown of the whole ion crystal. Actually, for the quasi-1D 218-ion crystal, the heating dynamics is too fast to be examined from the images. Therefore we first demonstrate the idea using a shorter chain of about $100$ ions whose heating dynamics is slower. As shown in Fig.~\ref{fig:cooling}(a)-(c), without the sympathetic cooling, the spots of individual ions quickly blur and mix up with each other. This process is fastest for the middle ions with small inter-ion distances, but even for ions on the edges it will finally become difficult to distinguish individual ions. On the other hand, with the sympathetic cooling beam turned on, the ions remain distinguishable and there is no visible change over $800\,$ms evolution time.

With this basic understanding about how the heating effects damage the qubit state detection, now we apply it to the 218-ion crystal. In Fig.~\ref{fig:cooling}(d) we pick up an ion (labeled as ion 1 above) and measure the photon counts in a $5\,$pixel$\times5\,$pixel box around it. With the sympathetic cooling beam turned off, the photon count quickly decays within $100\,$ms, and we estimate a decay time of $\tau=(256\pm14)\,$ms from the simplest fitting model of $e^{-t/\tau}$. This decay comes from two effects: as the ion heats up, it is more likely to move outside the selected region, and it scatters fewer photons due to the larger Doppler shifts. In the experiment, we use a threshold method to distinguish the bright ($|1\rangle$) and the dark ($|0\rangle$) state. Therefore the drop in the photon count of the bright state will significantly reduce the detection fidelity even before the timescale $\tau$, and is shorter than the coherence time $T_2$ measured before under sympathetic cooling.

Although we show that sympathetic cooling is critical for the long-time storage of the qubit states in the multi-ion crystal, it is also well-known that the scattered photons from the cooling ions can lead to crosstalk on the storage ions and thus limit the storage lifetime. This is because of the same transition frequency between the cooling ions and the storage ions. Previously, this is solved by using different ion species to encode the two qubit types \cite{wang2021single,PhysRevLett.118.053002,PhysRevA.65.040304,PhysRevA.68.042302,PhysRevA.79.050305}. In the future, when combined with the dual-type qubit scheme \cite{yang2022realizing} which encodes the data qubits and the ancilla qubits into different clock states of the same ion species, the crosstalk can be suppressed to allow smaller distance between the two qubit types and an enhanced storage lifetime, without the need to manipulate multiple ion species.

To sum up, we have demonstrated the multi-ion storage capacity with sub-second lifetime in a quasi-1D crystal of above 200 ions. This capability to coherently store a large number of qubits for long time is crucial for the quantum computing tasks in the future with deep circuit depth. It is also necessary for ion-photon quantum networks since the entanglement generation between different ion trap modules through photon links is typically much slower than the local gate operations inside individual modules \cite{hucul2015modular}.

\begin{acknowledgments}
This work was supported by the Innovation Program for Quantum Science and Technology (2021ZD0301601), the Tsinghua University Initiative Scientific Research Program, and the Ministry of Education of China. Y.-K. W. acknowledges in addition support from the start-up fund from Tsinghua University.
\end{acknowledgments}

%

\end{document}